\documentclass{article}
\usepackage{spconf,amsmath,graphicx}
\usepackage[urlcolor=blue]{hyperref}
\usepackage{booktabs}

\usepackage{lipsum}

\title{NeuFA: Neural network based end-to-end forced alignment with bidirectional attention mechanism}
\name{Jingbei Li$^1$, Yi Meng$^1$, Zhiyong Wu$^{1,2}$, %\thanks{* Corresponding author.},
Helen Meng$^{2}$, Qiao Tian$^3$, Yuping Wang$^3$, Yuxuan Wang$^3$}
\address{
    $^1$ %Tsinghua-CUHK Joint Research Center for Media Sciences, Technologies and Systems, \\
    Shenzhen International Graduate School, Tsinghua University, Shenzhen, China\\
    $^2$ %Department of Systems Engineering and Engineering Management, \\
         The Chinese University of Hong Kong, Hong Kong SAR, China\\
    $^3$ ByteDance, Shanghai, China\\
    \small{
        \{lijb19, my20\}@mails.tsinghua.edu.cn, 
        \{zywu, hmmeng\}@se.cuhk.edu.hk,
        \{tianqiao.wave, wangyuping, wangyuxuan.11\}@bytedance.com
    }
}
\begin{document}
\ninept
\maketitle
\begin{abstract}
Although deep learning and end-to-end models have been widely used
and shown their superiority 
in automatic speech recognition (ASR) and text-to-speech (TTS) synthesis,
state-of-the-art forced alignment (FA) models are still based on hidden Markov model (HMM).
HMM has limited view of contextual information and is developed with long pipelines,
leading to error accumulation and unsatisfactory performance.
Inspired by the capability of attention mechanism in capturing long term contextual information and learning alignments in ASR and TTS,
we propose a neural network based end-to-end forced aligner called NeuFA, 
in which a novel bidirectional attention mechanism plays an essential role.
NeuFA integrates the alignment learning of both ASR and TTS tasks 
in a unified framework
by learning bidirectional alignment information
from a
shared attention matrix in the proposed bidirectional attention mechanism.
Alignments are extracted from the
learnt attention weights and optimized by the ASR, TTS and FA tasks in a multi-task learning manner.
Experimental results demonstrate the effectiveness of our proposed model,
with mean absolute error (MAE) on test set drops from 25.8 ms to 23.7 ms at word level,
and from 18.0 ms to 15.7 ms at phoneme level
compared with state-of-the-art HMM based model.

\end{abstract}
\section{Introduction}
%\blfootnote{This work is partially supported by National Natural Science Foundation of China (NSFC) (62076144) and National Social Science Foundation of China (NSSF) (13\&ZD189).}
Forced alignment (FA),
which produces the bidirectional mapping between the given
text and speech sequences,
has been widely used in speech processing tasks over decades.
Such bidirectional mapping information 
provides the possibility of recognizing fine-grained prosody patterns of speakers,
from which high-level analysis
like sociolinguistical, phonetical and psycholinguistical analysis
could be built \cite{adda2011quantifying, dicanio2013using, labov2013one, schuppler2011acoustic, yuan2006towards}.

Conventionally, FA models are based on hidden Markov model (HMM)
\cite{rabiner_introduction_1986}, such as Prosodylab-Aligner \cite{gorman2011prosodylab} and Montreal forced aligner (MFA) \cite{mcauliffe_montreal_2017}.
HMM based models only have limited views of contextual information.
Moreover, these models need to be trained in long pipelines which 
not only
are 
resource-consuming,
but also lead to error accumulation and unsatisfactory results.

With the development of deep learning and end-to-end technology,
many neural network based models are proposed for automatic speech recognition (ASR) \cite{chan_listen_2016, gulati_conformer_2020} 
and text-to-speech (TTS) synthesis \cite{wang_tacotron:_2017, shen_natural_2017, ren_fastspeech_2019} and outperform conventional methods.
%The attention mechanism has also shown its capability of learning alignments in these tasks and whether the alignments are learnt
These models %are 
commonly %developed in encoder-decoder with attention framework \cite{bahdanau_neural_2014},
%from which the encoder and decoder convert the sequences between texts and speeches,
%in end-to-end manners and commonly employ 
%and 
employ the attention mechanism \cite{vaswani_attention_2017} to capture a long view of contextual information. %for each decoding step.
Besides, the attention mechanism used in ASR and TTS 
has also shown its capability of learning alignment information \cite{chan_listen_2016, wang_tacotron:_2017, shen_natural_2017} and improved model interpretability.
Then it is of possibility to build a neural network model for FA with the alignment information learnt by attention mechanism.
FastSpeech \cite{ren_fastspeech_2019} introduces a primary way to convert the alignment information to duration by %adding 
summing
the weights for each phoneme.
%Moreover, the alignment information can be learnt from both the ASR and TTS tasks.
Moreover, further improvement could be made if the alignments could be optimized by both ASR and TTS with multi-task learning.

%only the ASR direction is used to learn alignments in these models,
%and the TTS direction which is also known to be capable to learning alignments.
%and limited uses of the texts to be aligned since they are developed from ASR models.
%These lead to unsatisfied results in FA.
%than the end-to-end models used in ASR and TTS,
%Also the texts information are rarely considered when training these FA models 

In this paper,
to improve FA with deep learning and end-to-end technologies,
%inspired by the attention mechanism \cite{} and its capability of learning alignment information,
we propose a neural network based end-to-end forced aligner %which is called 
named NeuFA,
%in which a very essential component and algorithm is the novel bidirectional attention mechanism. 
in which the most essential algorithm is a novel bidirectional attention mechanism. 
%and bidirectional attention mechanism which is the most essential algorithm used in NeuFA.
The bidirectional attention mechanism are extended from the original attention mechanism to learn bidirectional relation mapping between two sets of key-value pairs.
In the neural network based FA task, the input includes both text and speech sequences, and the bidirectional alignment information between text and speech can be learned by two separately but related tasks from text to speech (TTS) and speech to text (ASR).
In this way, 
%And 
NeuFA integrates the alignment learning of both ASR and TTS tasks in a united framework by learning the bidirectional alignment information
%in the proposed bidirectional attention mechanism.
%The bidirectional alignment information is learnt by 
from
the shared attention matrix in bidirectional attention mechanism and optimized by the ASR, TTS and FA tasks with multi-task learning.
Then the attention weights for ASR and TTS tasks are obtained from the shared attention matrix
and converted to boundaries at word or phoneme level by a boundary detector.
%The bidirectional attention mechanism provides shared attention matrices for both the ASR and TTS directions for each pair of inputs.
%Then alignments can be extracted from the attention matrices and optimized by both directions in a multi-task learning manner.
%Then boundaries are extracted from these shared attention matrices at word and phoneme levels.
Experimental results demonstrate the effectiveness of our proposed model.
%,
%with mean absolute error (MAE) on test set drops from 25.8 ms to 23.7 ms at word level,
%and drops from 17.0 ms to 15.7 ms at phoneme level
%than MFA.
The source code of NeuFA is also available at https://github.com/thuhcsi/NeuFA.
%Then alignments are extracted from the 
%from which both the ASR and TTS tasks are con
%in which the input texts and speeches are trying to predict each other.
%And alignments are learnt though the shared attention matrices 
%in the proposed bidirectional attention mechanism
%used in both the ASR and TTS tasks
%for each pair of inputs.
%\input{content/figure2}
\section{Methodology}
\subsection{Bidirectional attention mechanism}
\label{bidirectional attention}
%In this section, we will firstly introduce bidirectional attention mechanism which is the most essential algorithm used in NeuFA,
%then the details of NeuFA which is our proposed neural network based FA model.

%Traditional attention mechanism is proposed and successfully applied in many researches.
%By giving a query and a set of key-value pairs,
%traditional attention mechanism produces a weighted sum of the values,
%where the weight assigned to each value is computed by a compatibility function of the query with the corresponding key.
%
%However, 
%the positions of the the query and keys 
%are not equivalent in traditional attention mechanism 
%since only the keys have corresponding values 
%and only the ``queries-to-keys" relations is learnt.
%Traditional attention mechanism are always be applied with parallel data such as the machine learning and FA corpora, 
%in which the queries and keys are always from two equivalent sequences.
%In such cases, bidirectional relations are of possible to be learnt 
%for both the ``queries-to-keys" and ``keys-to-queries'' directions
%at the same time if we have an attention mechanism with equivalent positions for queries and keys.

\begin{figure}[t]
	\centering
	\includegraphics[width=.8\linewidth]{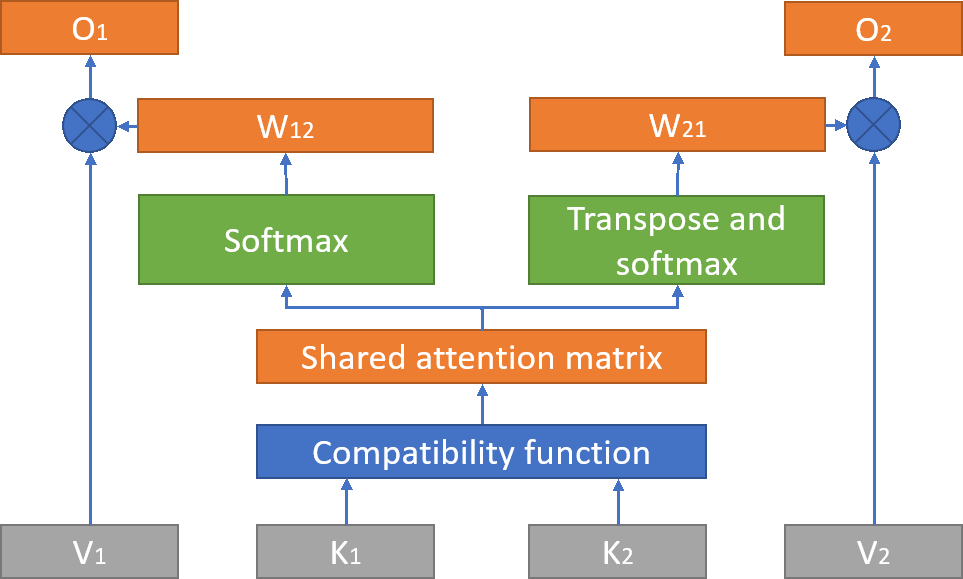}
	\caption{
	    Bidirectional attention mechanism
	}
	\label{fig:attention}
\end{figure}

To achieve bidirectional relation learning for parallel data,
we extend conventional attention mechanism \cite{vaswani_attention_2017} to bidirectional attention mechanism 
to learn the bidirectional relation mapping between two sets of key-value pairs, 
%which is 
as shown in Figure \ref{fig:attention}.
%By giving two sets of key-value pairs,
%the bidirectional attention mechanism produces two weighted sums of each sets of values,
%where the weights for each sets are computed from a shared compatibility function of the two sets of keys.
%This can also be formulated as:
%\begin{align}
%    O &= Attention(Q, K, V)
%\end{align}
%where $Q$, $K$ and $V$ are the input queries, keys and values,
%and $O$ is the weighted sum of the values.
%$n$ is the number of keys,
%$d_{q}$, $d_{k}$ and $d_{v}$ are the number of feature dimensions for query, keys and values,
%where $Q \in R^{d_q}$, $K \in R^{n \times d_k}$ and $V \in R^{n \times d_v}$ are the input query, keys and values,
%$n$ is the number of keys,
%$d_{q}$, $d_{k}$ and $d_{v}$ are the number of feature dimensions for query, keys and values,
%and $O \in R^{d_v}$ is the weighted sum of the values.
%from which 
%when the input queries and keys are 
%the text and speech sequences can be used as either the query sequence or keys.
%the query and keys are always from two sequences and these two sequences 
%For example,
%the text sequence is always used as the query sequence in ASR while it can also be used as the keys in TTS,
%and the spectrogram is always used as the query sequence in TTS while it can also be used as the keys in ASR.
%\subsubsection{Bidirectional multiplicative attention}
%\label{bidirectional multiplicative attention}
%The bidirectional attention mechanism takes two sets of key-value pairs as inputs,
Each set of keys serves as the queries for the other set of key-value pairs.
Then two weighted sums of the values are produced based on a shared compatibility function between the two sets of keys.
It is also formulated as:
\begin{align}
    A &= f(K_1, K_2) \\
    \label{bidirectional attention2}
    W_{12}, W_{21} &= softmax(A, A^T)\\
    O_1, O_2 &= W_{12}^{T}V_1, W_{21}^{T}V_2
\end{align}
%\begin{align}
%\label{bidirectional attention3}
%\end{align}
where $K_1 \in R^{n_1 \times d_{k1}}$, $V_1 \in R^{n_1 \times d_{v1}}$ and
$K_2 \in R^{n_2 \times d_{k2}}$, $V_2 \in R^{n_2 \times d_{v2}}$ 
are the two sets of key-value pairs,
$n_1$ and $n_2$ are the numbers of the key-value pairs, % in these two groups,
$d_{k1}$, $d_{v1}$, $d_{k2}$ and $d_{v2}$ are feature dimensions, %of keys and values in these two sets,
$f$ is the compatibility function,
$A \in R^{n_1 \times n_2}$ is the shared attention matrix,
$W_{12} \in R^{n_1 \times n_2}$ and
$W_{21} \in R^{n_2 \times n_1}$ 
are the attention weights for two directions respectively,
$O_1 \in R^{n_2 \times d_{v1}}$ is the weighted sum of $V_1$ for each key in $K_2$ and
$O_2 \in R^{n_1 \times d_{v2}}$ is the weighted sum of $V_2$ for each key in $K_1$.

Depending on how the compatibility function is implemented,
we propose a multiplicative form and an additive form for bidirectional attention mechanism.
For the multiplicative form (i.e. the bidirectional multiplicative attention),
the compatibility function are implemented as:
\begin{align}
\label{bidirectional multiplicative attention}
    A &= f_1(K_1) \times f_2(K_2)^T 
%    W_{21} &= softmax(S^T) \\
%    O_{2} &= W_{21}^{T}V_2
\end{align}
where
$f_1:R^{n_1 \times d_{k1}} \rightarrow R^{n_1 \times d_a}$ and
$f_2:R^{n_2 \times d_{k2}} \rightarrow R^{n_2 \times d_a}$ 
are linear projections,
$d_a$ is the 
%[zywu-1001]
%number of feature dimensions
feature dimension
of the attention hidden space.
%$f_s:R^{n_1 \times n_2 \times d_a} \rightarrow R^{n_1 \times n_2}$ 
%\begin{align}
%    S &= f_q(Q)^T \times f_k(K) \\
%    W &= softmax(S) \\
%    O &= W^{T}V
%\end{align}
%\subsubsection{Bidirectional additive attention}
The additive form or the bidirectional additive attention can also be deduced from additive attention \cite{bahdanau_neural_2014},
which is formulated as:
\begin{align}
    A &= f_a(dup_1(f_1(K_1)) + dup_2(f_2(K_2)))
\end{align}
where 
%are the two groups of key-value pairs,
%$f_1:R^{n_1 \times d_{k1}} \rightarrow R^{n_1 \times d_a}$,
%$f_2:R^{n_2 \times d_{k2}} \rightarrow R^{n_2 \times d_a}$ and
%$dup_1:R^{n_1 \times d_{k1}} \rightarrow R^{n_1 \times n_2 \times d_{k1}}$ and
%$dup_2:R^{n_2 \times d_{k2}} \rightarrow R^{n_1 \times n_2 \times d_{k2}}$ 
%duplicate the outputs of $f_1(K_1)$ and $f_2(K_2)$ to a same shape,
$dup_1:R^{n_1 \times d_{a}} \rightarrow R^{n_1 \times n_2 \times d_{a}}$ and
$dup_2:R^{n_2 \times d_{a}} \rightarrow R^{n_1 \times n_2 \times d_{a}}$ 
duplicate the outputs of $f_1(K_1)$ and $f_2(K_2)$ to a same shape,
$f_a:R^{n_1 \times n_2 \times d_a} \rightarrow R^{n_1 \times n_2}$ is a linear projection.

\subsection{Neural network based forced alignment}

\begin{figure}[t]
	\centering
	\includegraphics[width=.9\linewidth]{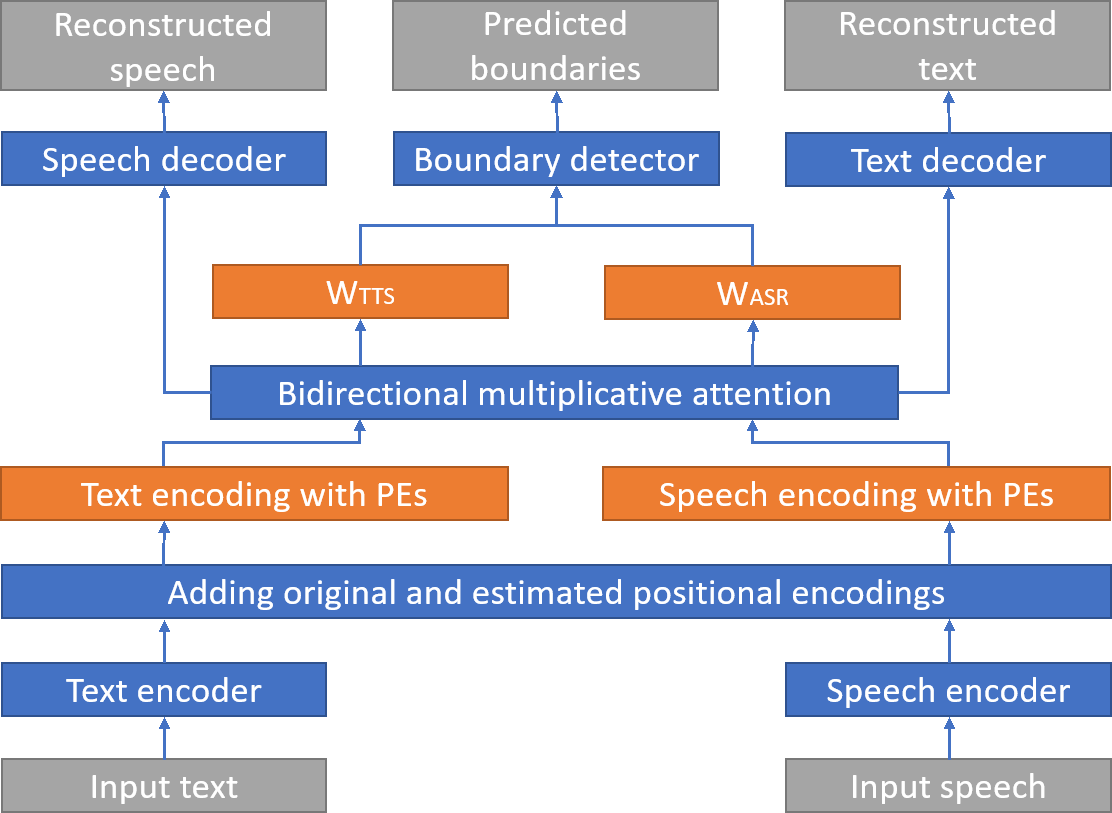}
	\caption{
	    Neural network based FA with bidirectional attention
	}
	\label{fig:neufa}
\end{figure}

In this section we will introduce our proposed neural network based FA model, NeuFA.
%consists of many modules and they are all essention.
As shown in Figure \ref{fig:neufa}, NeuFA has a text encoder and a speech decoder for the TTS task,
a speech encoder and a text decoder for the ASR task.
These two encoder-decoder modules share a same attention module implemented with the bidirectional attention mechanism introduced in Section \ref{bidirectional attention}.
Then a boundary detector is employed to extract the boundaries from the %alignment information 
%learnt %by the bidirectional 
attention weights in both ASR and TTS tasks.
%Estimated text and speech position encodings and diagonal attention loss are also proposed to help learning the alignments.

\subsubsection{Text encoder and speech encoder}
The text encoder and speech encoder convert the input text and speech sequences into text and speech encodings.
The encoder of Tacotron 2 \cite{shen_natural_2017} 
is employed as the text encoder in NeuFA since it
has already shown its ability in modeling text information in TTS.
The speech encoder is designed based on the content encoder \cite{qian_unsupervised_2020} proposed in voice conversion to capture the textual information in speeches. 
The speech encoder consists of 3 convolutional layers each containing 512 filters with
shape 17 $\times$ 1 and batch normalization, and two 256 dimensional bidirectional GRU \cite{cho_properties_2014} layers.
%The speech encoder converts the input speech sequences into speech encodings.

\begin{figure}[t]
	\centering
	\includegraphics[width=.9\linewidth]{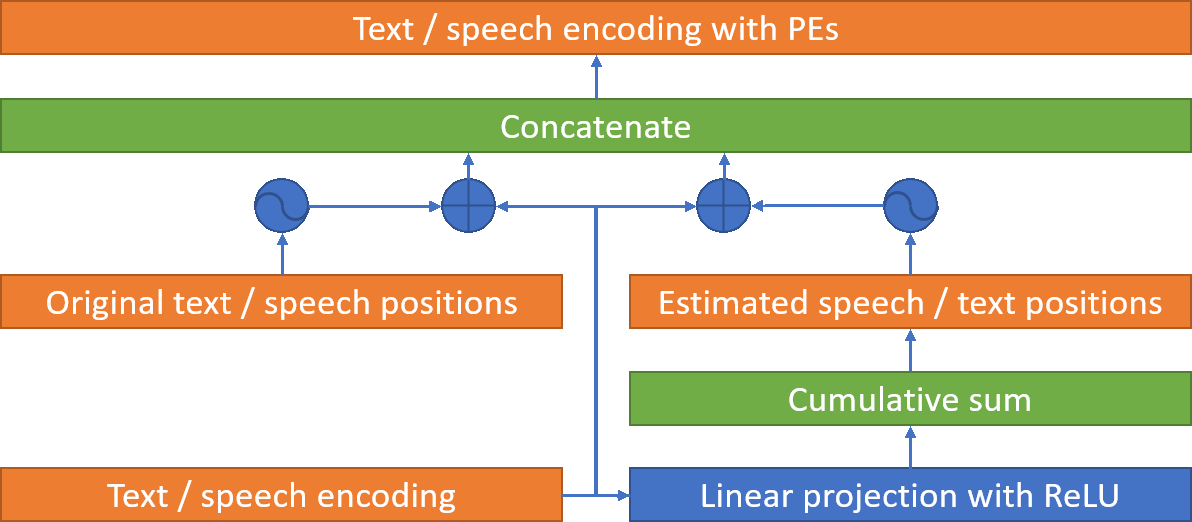}
	\caption{
	    Adding original and estimated positional encodings
	}
	\label{fig:pe}
\end{figure}

\subsubsection{Estimated positional encodings}

As shown in Figure \ref{fig:pe}, original and estimated positional encodings (PEs) are added to the text encodings and speech encodings before they are processed by the attention mechanism.

We employ the positional encodings in Transformer \cite{vaswani_attention_2017} as the original positional encodings for texts and speeches:
\begin{align}
\label{positional encoding1}
    PE_{(pos,2i)} &= sin(pos/10000^{2i/d})\\
\label{positional encoding2}
    PE_{(pos,2i+1)} &= cos(pos/10000^{2i/d})
\end{align}
where
$pos$ is the position,
$i$ is the index,
and $d$ is the number of the feature dimensions.

However, the positions (i.e. $pos$) are of different meanings in the original positional encodings for texts and speeches.
For texts, the positions are the indices of the input phonemes;
while for speeches, they are the indices of the frames of the input speeches.
Such mismatch in the positions may confuse the model to learn clear alignments.
Therefore, we propose estimated positional encodings 
%[zywu-1001]
%which are compatible with the original positional encodings
to alleviate this mismatch by estimating the possible speech or text positions from its corresponding text or speech encodings.

The estimated positional encodings are also computed with Equation \ref{positional encoding1} and \ref{positional encoding2},
except the positions are estimated positions rather than the indices of the input sequences.
For texts, estimated speech positions are computed from the text encodings.
And for speeches, estimated text positions are computed from the speech encodings.
In computing the positions,
the input encodings are first converted into estimated lengths by linear projection with $ReLU$ activation.
Then the estimated lengths are cumulatively summed into monotonic estimated positions.
The procedure can be formulated as:
\begin{align}
   pos'_s &= cumsum(ReLU(f_t(E_t)))\\
   pos'_t &= cumsum(ReLU(f_s(E_s)))
\end{align}
where $E_t$, $E_s$ are the input text and speech encodings,
$f_t$, $f_s$ are linear projections,
$pos'_s$, $pos'_t$ are the estimated cumulatively summed speech and text positions for the input text and speech sequences.

The original and estimated positional encodings are separately added to two copies of the input text and speech encodings, and then concatenated together:
\begin{align}
E'_t &= [E_t + PE_t; E_t + PE'_s]\\
E'_s &= [E_s + PE'_t; E_s + PE_s]
\end{align}
where
%$E_t$, $E_s$ are the input text and speech encodings,
$PE_t$, $PE_s$ are the original text and speech positional encodings,
$PE'_t$, $PE'_s$ are the estimated text and speech positional encodings,
$E'_t$, $E'_s$ are the output text and speech encodings with positional encodings added.

By introducing the estimated positional encodings to NeuFA, two additional losses called relative estimated text and speech length losses are used in back-propagation.
These two losses ensure that the estimated positions are compatible with the index based positions,
which can be formulated as:
\begin{align}
    loss_{l}^{t} &= MSE(1, last(pos'_t)/L_t) \\
    loss_{l}^{s} &= MSE(1, last(pos'_s)/L_s)
\end{align}
where
$L_t$, $L_s$ are the ground-truth lengths of the input text and speech sequences,
%$L'_t$, $L'_s$ are the sums of the estimated text and speech lengths which are also the last elements of $pos'_t$ and $pos'_s$,
$last$ returns the last elements of $pos'_t$ and $pos'_s$ which correspond to the total estimated text and speech lengths,
$MSE$ is mean squared error.
%[zywu-1001]前面已经说了
%$loss_{l}^{t}$ and $loss_{l}^{s}$ are the relative estimated text and speech length losses.

In practise, we find that the estimated text positional encodings are significantly helpful to learn clear alignments than just original positional encodings.
Moreover, these positional encodings are vital to distinguish those phonemes with the same pronunciation in sentences.
%Ablation studies about these positional encodings will be discussed in Section \ref{ablation}.

\subsubsection{Bidirectional attention mechanism}
As shown in Figure \ref{fig:neufa}, 
the bidirectional attention takes the text and speech encodings with 
%positional encodings 
PEs
as inputs,
and respectively summarizes textual and acoustic information for the 
%following 
TTS and ASR tasks.
%The attention weights used to summarize the text and acoustic information are shared between the two tasks,
%and from which the alignments are learnt in a multi-task learning manner.
We use an 128 dimensional bidirectional multiplicative attention %as described in Equation \ref{bidirectional multiplicative attention}
for NeuFA since it is more efficient in memory usage.

\subsubsection{Text decoder and speech decoder}
The text decoder and speech decoder respectively reconstruct the input text and speech sequences from the summarized acoustic and textual information.
The text decoder consists of a stack of two 128 dimensional bidirectional LSTM layers and a linear projection with softmax activation.
The speech decoder has a similar structure, consisting of a stack of two 256 dimensional bidirectional LSTM layers and a linear projection.
Cross entropy and MSE are used as the loss functions for the reconstructed text and speech sequences:
%, which are defined as:
\begin{align}
loss_{t} &= CrossEntropy(T', T) \\
loss_{s} &= MSE(S', S)
\end{align}
where $T$ and $S$ are the input text and speech sequences,
$T'$ and $S'$ are the reconstructed text and speech sequences.

\subsubsection{Boundary detector}
Instead of directly predicting the boundaries,
the boundary detector takes the attention weights from both ASR and TTS directions as inputs
%consists of a stack two 128 dimensional bidirectional LSTM layers and a linear projection with ReLU activation
to predict %the the left and right boundaries for each phoneme.
the left and right boundary signals for each phoneme.
The boundary signals are inspired by the stop token in Tacotron \cite{wang_tacotron:_2017} and Tacotron 2 \cite{shen_natural_2017},
which can provide more fine-grained gradients for each position than just the values of boundaries.
Ground-truth boundary signals are generated from the annotated boundaries,
in which the elements are set to 0 if they are before the boundaries and 1 otherwise.
Therefore the predicted boundary signals are also monotonic signals ranged from 0 to 1.
During inference, the positions of the first elements whose boundary signals are greater than 0.5 are used as the predicted boundaries. 
%The elements near the diagonal of the  are used as the inputs 

A 6 channel feature matrix is generated from the input attention weights,
including the original attention weights
and their cumulative sums in both forward and backward directions:%, %along the temporal dimension,
%which can be formulated as:
\begin{align}
    \nonumber
    F = [&W_{TTS},  cumsum(W_{TTS}),  r(cumsum(r(W_{TTS}))), \\
        &W_{ASR}^{T}, cumsum(W_{ASR}^{T}), r(cumsum(r(W_{ASR}^{T})))]
\end{align}
where $W_{TTS}$ and $W_{ASR}$ are the attention weights for the TTS and ASR directions,
$r$ reverses its input along the temporal dimension and $F$ is the output 6 channel feature matrix.
$W_{ASR}$ is transposed
%in these equations 
due to the Equation \ref{bidirectional attention2} in bidirectional attention mechanism.

The feature matrix $F$ is then processed by
a stack of 3 convolutional layers each containing 32 filters with shape 17 $\times$ 17
and a linear projection followed by the $sigmoid$ activation to detect the left and right boundaries.
The outputs are cumulatively summed and converted to monotonic boundary signals ranged from 0 to 1 by the $tanh$ activation for each left and right boundaries.
%This procedure can be formulated as:
\begin{align}
    B' = tanh(cumsum(sigmoid(f(convs(F)))))
\end{align}
where 
$convs$ is the stacked convolutions,
$f$ is linear projection and
$B'$ is the predicted boundary signals.

\begin{figure}[t]
	\centering
	\begin{minipage}[b]{0.49\linewidth}
		\centering
		\includegraphics[width=.85\linewidth]{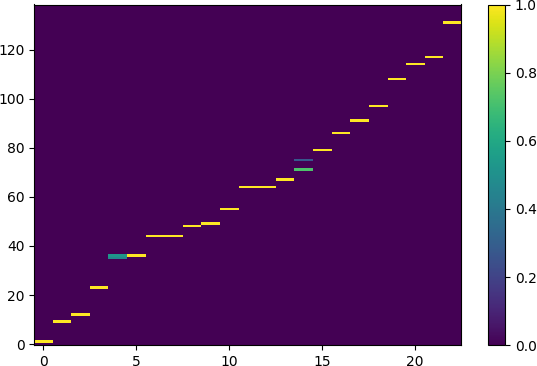}\\
		(a) Attention weights for ASR\\
		\includegraphics[width=.85\linewidth]{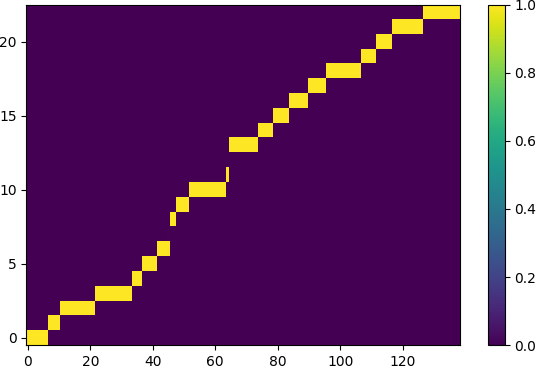}\\
		(b) Attention weights for TTS
	\end{minipage}
	\hfill
	\begin{minipage}[b]{0.49\linewidth}
		\centering
		\includegraphics[width=.85\linewidth]{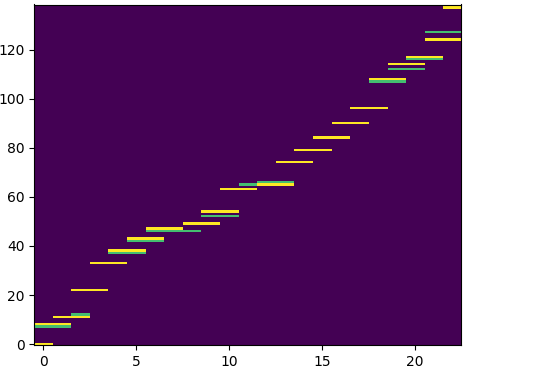}\\
		(c) Predicted boundaries\\
		\includegraphics[width=.85\linewidth]{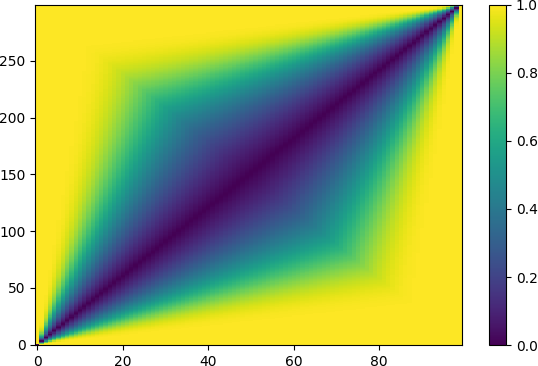}\\
		(d) Diagonal constraint matrix
	\end{minipage}
	\caption{
	    (a) Learnt attention weights for ASR.
	    (b) Learnt attention weights for TTS.
	    (c) Illustration of predicted (green) and ground-truth (yellow) phoneme boundaries.
	    (d) Illustration of a diagonal constraint matrix used in the diagonal attention loss.
    }
	\label{fig:attention weights}
\end{figure}

MAE of the predicted and ground-truth boundary signals is used as the loss function for boundary detector:
\begin{equation}
    loss_{b} = MAE(B', B)
\end{equation}
where $B$ and $B'$ are the ground-truth and predicted boundary signals respectively.

\subsubsection{Diagonal attention loss}

Diagonal attention loss is proposed to help the model learn alignments
by multiplying the elements in attention weight matrices with diagonal constraints.
The diagonal attention loss is defined as:
\begin{align}
    loss_{a} &= mean\left((W_{TTS} + W_{ASR}^T) \odot D\right) \\
    \label{diagnol}
    D_{i,j} &= tanh\left(\frac{1}{2}max\left(\frac{p}{q}, \frac{q}{p}, \frac{1-p}{1-q}, \frac{1-q}{1-p}\right)\right)
\end{align}
where
$\odot$ is element-wise multiplication
and 
$D$ %is the matrix of diagonal weights,
%Diagonal weights are 
is dynamically generated %for each input attention matrix 
with Equation \ref{diagnol}
where
$i$, $j$ are the indices,
$p=\frac{i}{n_1}$, $q=\frac{j}{n_2}$ are the relative position of $i$, $j$ in their corresponding dimensions.
An illustration of $D$ is also shown in Figure \ref{fig:attention weights} (d).

The final loss function for training NeuFA is then defined as:
\begin{equation}
    loss = \alpha loss_{t} + \beta loss_{s} + \gamma loss_{l}^{t} + \delta loss_{l}^{s} + \epsilon loss_{a} + \zeta loss_{b}
\end{equation}
where $\alpha$ to $\zeta$ are loss weights.

%In practise, the proposed diagonal attention loss is vital to learn alignments.

\section{Experiments}
\subsection{Training setups}
We follow the training setups of the MFA \cite{mcauliffe_montreal_2017} to train NeuFA except the MFCCs are extracted with librosa \cite{mcfee2015librosa} rather than Kaldi \cite{povey_kaldi_2011},
and graphme-to-phoneme conversions are made by a pretrained model \cite{li_pre-trained_2021}.
We train an NeuFA for word level as if only orthographic transcription in Buckeye \cite{pitt_buckeye_2005} were known,
and another NeuFA for phoneme level with the phoneme sequences given in Buckeye.

Each NeuFA model is firstly trained on the full set of the LibriSpeech \cite{panayotov_librispeech_2015} corpus for 120,000 steps with a batch size of 16.
The learning rate is fixed to $10^{-4}$.
Loss weights $\alpha$ to $\delta$ are simply set to 0.1, 1, 10, 10 to balance the losses to a same level.
$\epsilon$ is set to 1,000 to help the model learning alignments.
%Loss weights $\alpha$ to $\epsilon$ are simply set to 0.1, 1, 10, 10 and 1000 to balance the losses to a same level.
$\zeta$ is set to 0 since there is no boundary annotation for the LibriSpeech corpus.

Then each model is trained on the data of 36 speakers in the Buckeye corpus for 220,000 steps with a batch size of 16 to train the boundary detector.
The learning rate is also fixed to $10^{-4}$.
Loss weights $\alpha$ to $\delta$ are same to those in the previous stage.
$\zeta$ is now set to 100 and $\epsilon$ is set to 0 since the boundary loss can provide more accurate alignment information than the diagonal attention loss. 

The experiments are implemented with PyTorch \cite{paszke_pytorch_2019} on an NVIDIA Tesla V100 GPU.

\subsection{Experimental results}

The data of the rest 4 unseen speakers in the Buckeye corpus are used as the test set.
The models are evaluated on the MAE and medians of absolute errors of the predicted left and right boundaries at word and phoneme levels.
Then the comparisons are made with the model shipped in the MFA package as the baseline.
\begin{table}[t]
  \caption{Performances of the baseline and proposed approaches}
  \label{table1}
  \centering
  \begin{tabular}{lcccc}
    \toprule
    & \multicolumn{2}{c}{\textbf{Word level}} &\multicolumn{2}{c}{\textbf{Phoneme level}} \\
    \textbf{Approach}&\textbf{mean} &\textbf{median} &\textbf{mean} &\textbf{median} \\
    \midrule
    \textbf{MFA} & 25.8 ms & 12.3 ms & 18.0 ms & 10.0 ms\\
    \textbf{NeuFA} & \textbf{23.7} ms & \textbf{9.0} ms & \textbf{15.7} ms & \textbf{9.1} ms\\
    - w/o EPEs & 32.1 ms & 11.5 ms & 19.5 ms & 9.9 ms \\
    - w/o TPEs & 33.8 ms & 12.0 ms & 20.7 ms & 10.0 ms \\
    - w/o SPEs & 24.2 ms & 9.1 ms & 16.8 ms & 9.2 ms \\
    - w/o ASR & 37.8 ms & 14.0 ms & 24.6 ms & 10.8 ms \\
    - w/o TTS & 50.7 ms & 18.7 ms & 33.5 ms & 15.8 ms \\
    - w/o DAL & 26.6 ms & 10.0 ms & 18.7 ms & 10.1 ms\\
    \bottomrule
  \end{tabular}
\end{table}
\begin{table}[htb]
  \caption{Accuracies at different tolerances for different approaches}
  \label{table2}
  \centering
  \begin{tabular}{lcccc}
    \toprule
    %& \multicolumn{2}{c}{\textbf{Word level}} &\multicolumn{2}{c}{\textbf{Phoneme level}} \\
    \textbf{Approach}&\textbf{10 ms} &\textbf{25 ms} &\textbf{50 ms} &\textbf{100 ms} \\
    \midrule
    \textbf{MFA (word)} & 0.41 & 0.78 & 0.91 & 0.96 \\
    \textbf{NeuFA (word)} & \textbf{0.55} & \textbf{0.82} & \textbf{0.92} & \textbf{0.96} \\
    \textbf{MFA (phoneme)} & 0.50 & 0.84 & 0.94 & 0.98 \\
    \textbf{NeuFA (phoneme)} & \textbf{0.55} & \textbf{0.87} & \textbf{0.95} & \textbf{0.98} \\
    \bottomrule
  \end{tabular}
\end{table}

As shown in Table \ref{table1}, 
NeuFA outperforms MFA at both word and phoneme levels.
The MAE is reduced to 23.7 ms and 15.7 ms respectively.
And the medians drops to 9.0 ms and 9.1 ms,
which means that NeuFA always predicts more accurate boundaries than MFA.
This can also be demostrated by the accuracies at different tolerances (percentage below a cutoff) evaluated and shown in Table \ref{table2}.
The accuracies are improved by 
0.14, 0.04, 0.01 at 10, 25 and 50 ms tolerances for word level,
and
0.05, 0.03, 0.01 at corresponding tolerances for phoneme level.
At 100 ms tolerance, the performance of NeuFA is on par with MFA.
\subsection{Ablation studies}
\label{ablation}
%We conduct ablation studies to 
%evaluate
%the effectiveness of different modules in NeuFA,
%and their results are also 
%shown in Table \ref{table1}.

In ablation studies,
the performance drops a lot
if we remove the estimated postional encodings (EPEs) and only use the original positional encodings to train the model.
This illustrates the mismatch between original text and speech position encodings.
%This demonstrates our hypothesis about the mismatch between the original text and speech position encodings.

The performance also drops if we remove the original and estimated text positional encodings (TPEs) and only use the original and estimated speech positional encodings (SPEs) to train the model.
However, it barely drops if we do the opposite.
This is reasonable since estimating the number of phonemes from speech is more easier and accurate than estimating the duration of speech from phonemes. %sequences.

Significant performance loss will occur 
%The MAE raises to 45.1 ms and 54.7 ms 
if the ASR or TTS task is removed by setting the corresponding loss weight to 0.
This shows the necessity of both the ASR and TTS tasks for FA, and also the necessity of the proposed bidirectional attention mechanism.

%And ASR models may be insufficient for FA (e.g. you don't have to listen the whole speech of a phoneme to recognize it).
No alignment is learnt on the LibriSpeech corpus if we remove the diagnol attention loss (DAL) by setting $\epsilon$ to 0.
The training on the Buckeye corpus will then be similar to the training from random initialization.
Although the model can still learn alignments with the full architecture of NeuFA, 
it suffers from the absence of fine-trained bidirectional text-speech mapping information. %and having loss in performance.
%and the MAE raises to 26.6 ms.
\section{Conclusions}

To improve FA with deep learning and end-to-end technologies,
%and take full use of both the acoustic and textual information in inputs,
%and simply the development of FA models 
%in this paper,
we propose a neural network based end-to-end forced aligner named NeuFA
based on the bidirectional attention mechanism which is also proposed in this paper.
NeuFA integrates the alignment learning of both ASR and TTS in a united framework by the proposed bidirectional attention mechanism.
Attention weights for two tasks are obtained from the shared attention matrix in bidirectional attention mechanism,
and converted to boundaries at word or phoneme level by a boundary detector.
The effectiveness of NeuFA has been demonstrated by experimental results and ablation studies.
%with both the MAE at word and phoneme levels outperforming state-of-the-art HMM based model.
%with both the MAE at word and phoneme levels outperforming state-of-the-art HMM based model.

\textbf{Acknowledgment}: This work is partially supported by National Natural Science Foundation of China (NSFC) (62076144) and National Social Science Foundation of China (NSSF) (13\&ZD189).
%\blfootnote{\textbf{Acknowledgment:} This work is partially supported by National Natural Science Foundation of China (NSFC) (62076144) and National Social Science Foundation of China (NSSF) (13\&ZD189).}

%The bidirectional attention mechanism provides shared attention matrices for both the ASR and TTS directions for each pair of inputs.
%Then alignments can be extracted from the attention matrices and optimized by both directions in a multi-task learning manner.
%Then boundaries are extracted from these shared attention matrices at word and phoneme levels.
%with mean absolute error (MAE) of word boundaries dropped from 26 ms to 24 ms 
%and MAE of phoneme boundaries dropped from 17 ms to 15 ms than 
%Then alignments are extracted from the 
%from which both the ASR and TTS tasks are con
%in which the input texts and speeches are trying to predict each other.
%And alignments are learnt though the shared attention matrices 
%in the proposed bidirectional attention mechanism
%used in both the ASR and TTS tasks
%for each pair of inputs.

\bibliographystyle{IEEEbib}
\bibliography{references}

%\newpage
%\input{template/template}

\end{document}